\documentclass[
prapplied,
longbibliography,
reprint,
superscriptaddress,
floatfix,
citeautoscript,
amsmath,
amssymb
]{revtex4-2}

\usepackage[utf8]{inputenc}
\usepackage[T1]{fontenc}  
\usepackage{graphicx}
\usepackage{dcolumn}
\usepackage[separate-uncertainty]{siunitx}
\usepackage{booktabs}
\usepackage[dvipsnames]{xcolor}
\usepackage{comment}
\usepackage[colorlinks = true, urlcolor = blue, linkcolor = blue, citecolor = blue]{hyperref}

\begin{document}
\newcolumntype{C}[1]{>{\centering\let\newline\\\arraybackslash\hspace{0pt}}m{#1}}

\title{Impact of biased cooling on the operation of undoped silicon quantum well field-effect devices for quantum circuit applications}

\author{Laura K. Diebel}
\author{Lukas G. Zinkl}
\author{Andreas H\"otzinger}
\affiliation{Fakult\"at f\"ur Physik, Universit\"at Regensburg, 93040~Regensburg, Germany}
\author{Felix Reichmann}
\affiliation{IHP – Leibniz-Institut für innovative Mikroelektronik, 15236~Frankfurt~(Oder), Germany}
\author{Marco Lisker}
\affiliation{IHP – Leibniz-Institut für innovative Mikroelektronik, 15236~Frankfurt~(Oder), Germany}
\author{Yuji Yamamoto}
\affiliation{IHP – Leibniz-Institut für innovative Mikroelektronik, 15236~Frankfurt~(Oder), Germany}
\author{Dominique Bougeard}
\email{dominique.bougeard@ur.de}
\affiliation{Fakult\"at f\"ur Physik, Universit\"at Regensburg, 93040~Regensburg, Germany}

\date{\today}

\begin{abstract}
Gate-tunable semiconductor nanosystems are getting more and more important in the realization of quantum circuits. While such devices are typically cooled to operation temperature with zero bias applied to the gate, \textit{biased cooling} corresponds to a non-zero gate voltage being applied before reaching the operation temperature. We systematically study the effect of biased cooling on different undoped SiGe/Si/SiGe quantum well field-effect stacks~(FESs), designed to accumulate and density-tune two-dimensional electron gases (2DEGs). In an empirical model, we show that biased cooling of the undoped FES induces a static electric field, which is constant at operation temperature and superimposes onto the field exerted by the top gate onto the 2DEG. We show that the voltage operation window of the field-effect-tuned 2DEG can be chosen in a wide range of voltages via the choice of the biased cooling voltage. Importantly, quality features of the 2DEG such as the mobility or the temporal stability of the 2DEG density remain unaltered under biased cooling. We discuss how this additional degree of freedom in the tunability of FESs may be relevant for the operation of quantum circuits, in particular for the electrostatic control of spin qubits.
\end{abstract}

\maketitle

\section{Introduction}
\label{sec:Intro}
Field-effect devices are an important building block for the realization of quantum circuits in semiconductor heterostructures \cite{Hanson.2007,Manchon.2015,Scappucci.2021,Flensberg.2021,Place.2021,Thureja.2022}, in particular since they offer a gate-tunability of electric carriers, down to the nanoscale. It becomes more and more clear that a precise understanding of the electrostatics created by gate tuning in semiconductor heterostructure field-effect stacks~(FESs) are highly relevant for the stable operation of quantum circuits \cite{Connors.2019,Struck.2020,Kranz.2020,DegliEsposti.2022,Wuetz.2023,Meyer.2023,Massai.2023, Prager.2021}. For cryogenic operation of such devices, the potential applied to a gate during the cool-down process of the device from room temperature is of importance and may represent a tuning parameter of the electrostatics. Typically, quantum circuits are cooled down with zero bias applied to the gates. \textit{Biased cooling} represents the cool-down under a non-zero applied gate voltage. Biased cool-down has been studied for modulation-doped GaAs/AlGaAs quantum well (QW) heterojunctions in the context of the operation of two-dimensional electron gases (2DEGs) \cite{Long.1993,Buks.1994,Buks.1994b,Coleridge.1997} and of quantum point contacts \cite{PioroLadriere.2005,Buizert.2008}. In these works, the observed impact on the operation of the devices have been phenomenologically linked to the presence of dopant-induced defects and to leakage of Schottky gates. The statistical nature of dopant-induced defects and of the presence of leakage has limited the application of biased cooling as an additional degree of freedom for the device operation. More recently, in particular for spin qubit quantum circuits, FES based on undoped semiconductor heterostructures and including oxide-based dielectrics instead of Schottky gates are used \cite{Maune.2012,Prager.2021,Kammerloher.2021}. The absence of dopant-induced defects and the dielectric/semiconductor interface in the FES create an electrostatic environment in which biased cooling has not been considered yet.

In this paper, we systematically study the effect of biased cooling on different undoped SiGe/Si/SiGe QW FESs. We show that biased cooling with a voltage $U_{\text{BC}}$ induces a static electric field within the FES. At the operation temperature of the device, here \SI{1.5}{\kelvin}, this static electric field is insensitive to variations of the top gate voltage $U_{\text{TG}}$ and overlays the action of $U_{\text{TG}}$ on the 2DEG. As a result, the accumulation voltage of the 2DEG, as well as the whole field-effect tuning range of the 2DEG density $n_{\text{e}}$ can be deterministically shifted to a chosen voltage range by the appropriate choice of $U_{\text{BC}}$. Notably, shifting the 2DEG operation range does not impact 2DEG quality markers such as the electron mobility, temporal stability of the 2DEG density or the minimal detectable 2DEG density. In an empirical model, we show that the charge density which induces the static electric field is localized at the dielectric/heterostructure interface of the FES. These observations remain valid for FES with and without an oxidized Si cap at the interface to the dielectric.

\section{Field-effect stacks and experimental setup}
\label{sec:Material_Methods}

\begin{figure}[h]
    \includegraphics[width=\linewidth]{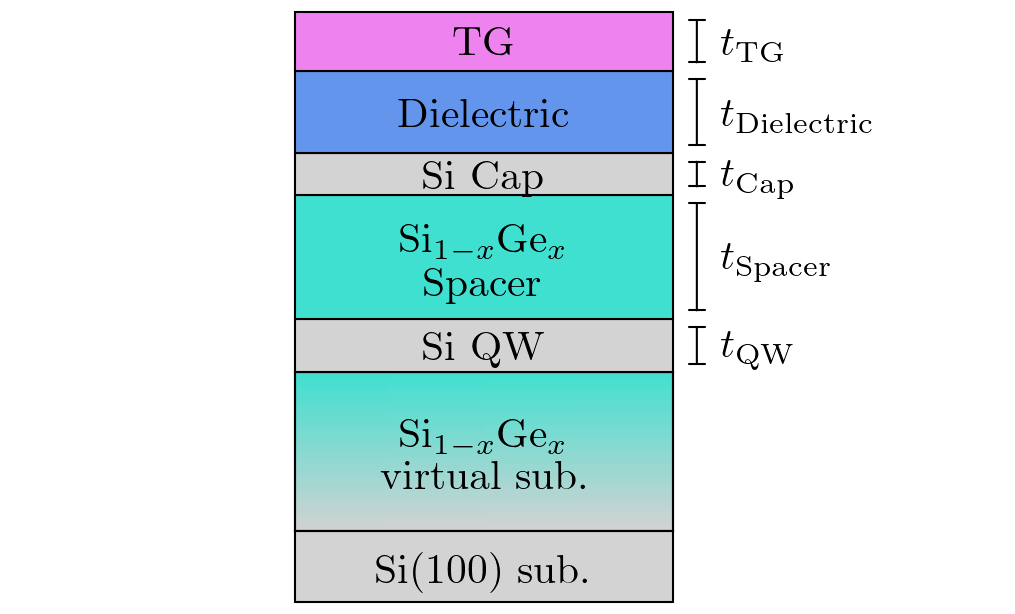}
    \caption{\label{fig:heterostructures} Cross-section schematic of the undoped Si/SiGe FESs. Details on the investigated FES~A, B and C are listed in Tab.~\ref{tab:FES}.}
\end{figure}

We investigated the biased cooling effect on three different undoped Si/SiGe FESs capable of hosting a 2DEG. These FESs are standard undoped Si$_{1-x}$Ge$_x$/Si/Si$_{1-x}$Ge$_x$ quantum well heterostructures for field-effect applications \cite{Lu.2009, Maune.2012, Neul.2024}. All of them share the same functional layer structure shown in Fig.~\ref{fig:heterostructures}, but with some crucial differences summarized in Tab.~\ref{tab:FES}. One of the key differences to highlight is the absence of a (oxidized) Si cap on FES~A. FES~A was grown and fabricated in a semi-industrial environment whereas the FESs~B and C were both fabricated in an academic clean room. The characterization of the FESs was done by Hall-bar geometry magneto-transport measurements at a temperature of \SI{1.5}{\kelvin} using standard lock-in techniques.
As an important reference point during our investigation we define the accumulation point with the corresponding accumulation voltage $U_{\text{Acc}}$. This denotes the voltage applied at the topgate~(TG), named $U_{\text{TG}}$, that is necessary to accumulate a 2DEG within our QW. We verified via quantum Hall signatures that the 2DEGs are accumulated beyond the metal-insulator transition when the current through the Hall-bar channel exceeds \SI{48}{\nano\ampere} of the \SI{50}{\nano \ampere} applied by the lock-in.

\begingroup
\squeezetable
\begin{table}[h]
\begin{ruledtabular}
\begin{tabular}{lccc}
FES & A & B & C\\
\midrule
Dielectric (method) & SiOx (HDP) & AlOx (ALD) & AlOx (ALD)\\
TG & TiN & Ti/Au & Ti/Au\\
$t_{\text{TG}}$ (\SI{}{\nano\meter}) & 30 & 10/100 & 10/100\\
$t_{\text{Dielectric}}$ (\SI{}{\nano\meter}) & 10 & 20 & 50\\
$t_{\text{Cap}}$ (\SI{}{\nano\meter}) & 0 & 1.5 & 1.5\\
$t_{\text{Spacer}}$ (\SI{}{\nano\meter}) & 33 & 45 & 45 \\
$t_{\text{QW}}$ (\SI{}{\nano\meter}) & 7 & 12 & 12\\
$x$ & 0.34 & 0.26 & 0.32\\
Heterostructure growth & CVD & MBE & MBE\\
Fabrication & semi-industrial & academic & academic\\
\end{tabular}
\end{ruledtabular}
\caption{\label{tab:FES} Overview of the three undoped SiGe/Si/SiGe quantum well FES~A, B and C. The abbreviations are introduced in Fig.~\ref{fig:heterostructures}.}
\end{table}
\endgroup

\section{Results}
\label{sec:Experimental_results}

\subsection{Impact of biased cooling on the 2DEG characteristics}

\begin{figure}
    \includegraphics[width=\linewidth]{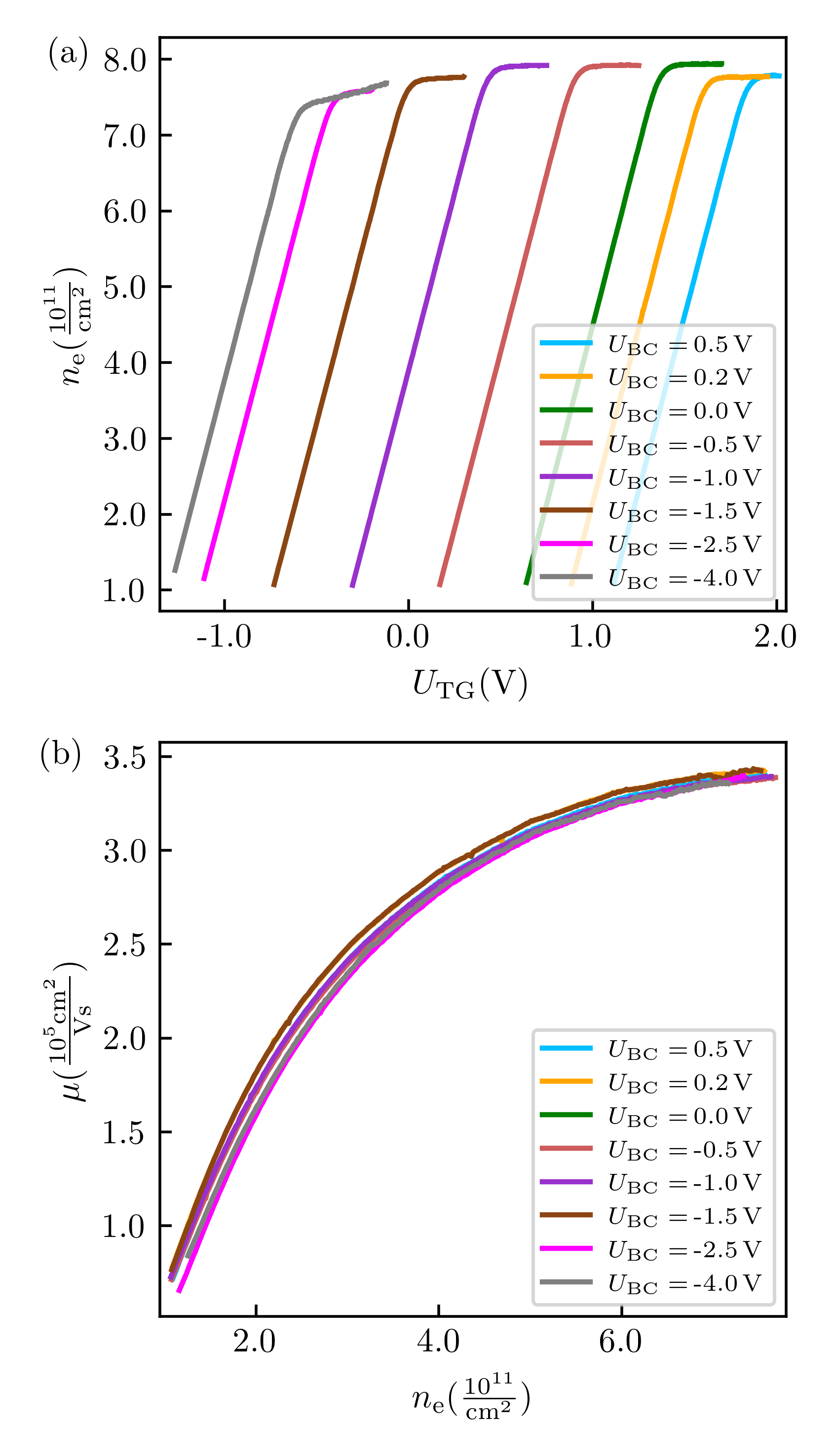}
    \caption{\label{fig:BC_shift} Impact of biased cooling on the electron density and the mobility of the 2DEG, representatively shown for FES~A at \SI{1.5}{\kelvin}. The definitions of the different voltages and of $n_{\text{e,max}}$ are found in the text.
    (a)~2DEG density~$n_{\text{e}}$ as a function of $U_{\text{TG}}$ for different biased cooling voltages $U_{\text{BC}}$.
    (b)~2DEG mobility~$\mu$ dependence on $n_{\text{e}}$ within the linear capacitive coupling regime ($n_{\text{e}} < n_{\text{e,max}}$) for different biased cooling voltages $U_{\text{BC}}$.}
\end{figure}

We find the heterostructures in all three FESs to be conducting at room temperature for any $U_{\text{TG}}$ value, even at $U_{\text{TG}} =$~\SI{0}{\volt}. At the contrary, we observe the conductance of the heterostructure to freeze-out during the cool-down to \SI{1.5}{\kelvin}, stating that the 2DEGs are normally off at $U_{\text{TG}} =$~\SI{0}{\volt}. To explore the influence of the biased cooling on the transport properties of a 2DEG, we apply a non-zero voltage at the TG while cooling down the field-effect stacks from room temperature to \SI{1.5}{\kelvin}. We refer to this voltage applied during the cool-down as \textit{biased cooling voltage} $U_{\text{BC}}$. We cooled down the FESs various times with varying $U_{\text{BC}}$ and determined the electron density as a function of the applied TG voltage ($U_{\text{TG}}$ sweep) at \SI{1.5}{\kelvin} for each cool-down. These measurements were performed for all three FES~A, B and C. Figure~\ref{fig:BC_shift}\,(a) representatively shows the results for FES~A. The FES cooled down with the commonly used $U_{\text{BC}} =$~\SI{0}{\volt} shows the state-of-the-art behavior of a 2DEG accumulating within the Si QW at a positive $U_{\text{TG}}$. In the classical approximation of a capacitor where the 2DEG and the TG are the plates, the electron density $n_{\text{e}}$ of the 2DEG within the QW depends linearly on the voltage $U_{\text{TG}}$ applied at the TG. This approximation starts to break down as soon as the potential difference is strong enough for electrons to tunnel out of the QW, through the SiGe barrier, into the interface between the heterostructure and the dielectric. In this saturation regime, the $n_{\text{e}}$ does not further increase with $U_{\text{TG}}$ \cite{Lu.2011,Wild.2012,Huang.2014,Laroche.2015}. As seen in Fig.~\ref{fig:BC_shift}\,(a), we find this characteristic behavior of a linear increase of $n_{\text{e}}$, followed by a saturation, to be independent of the applied $U_{\text{BC}}$. In quantum Hall experiments for selected negative and positive $U_{\text{BC}}$, we have verified that the electron density contributing to the transport after cooling down with a non-zero $U_{\text{BC}}$ is exclusively located in the QW 2DEG, excluding parallel conductance. Comparing the electron density curves, we observe a shift induced by $U_{\text{BC}}$. For positive biased cooling voltages applied during the cool-down, the electron density curves shift towards more positive/higher $U_{\text{TG}}$, whereas for negative biased cooling they shift towards more negative/lower $U_{\text{TG}}$. The shift increases with the absolute value of the $U_{\text{BC}}$ applied during cool-down.

In the remainder of this paper, we focus on the linear electron density tuning regime, which is the operation region for most field-effect device applications, including quantum circuits. We have verified that the 2DEG densities $n_{\text{e}}$ are reproducible within cool-downs in all three FESs (no hysteresis of the electron density in a $U_{\text{TG}}$ sweep). They are also reproducible among separate cool-downs for a given $U_{\text{BC}}$ value. Analyzing the minimum electron density at the accumulation point for the different $U_{\text{BC}}$, we see an average value of the minimum electron density of $n_{\text{e,min}} =$~\SI{1.1e11}{1\per\centi\square\meter} for FES~A in Fig.~\ref{fig:BC_shift}\,(a). We defined the maximum electron density~$n_{\text{e,max}}$ of the linear regime as the end of this range by extracting the first $n_{\text{e}}$ value which deviates by 2\% from a linear fit of the data. For FES~A we evaluate the average value of the maximum electron density to be $n_{\text{e,max}} =$~\SI{7.5e11}{1\per\centi\square\meter}. Across all FES, we observe no systematic dependence of the minimum electron density~$n_{\text{e,min}}$ as well as of the maximum electron density~$n_{\text{e,max}}$ on $U_{\text{BC}}$. Also, the slope of the 2DEG density's $U_{\text{TG}}$-dependence - which represents the capacitive coupling between the TG and the 2DEG - is unaffected by biased cooling with $U_{\text{BC}} \neq $~\SI{0}{\volt} for all three FESs (see Fig.~\ref{fig:BC_shift}\,(a) representatively for FES~A). In Fig.~\ref{fig:BC_shift}\,(b), we report the 2DEG mobility~$\mu$ as a function of $n_{\text{e}}$ for all tested $U_{\text{BC}}$, representatively for FES~A. No impact of $U_{\text{BC}}$ on the $\mu$ is observed. Note that remote scatterers should particularly manifest in the region of steep mobility increase below circa $n_{\text{e}} <$~\SI{4e11}{1\per\centi\square\meter} in FES~A, while scatterers in the QW will dominate beyond. \cite{Mi.2015,Laroche.2015}

To summarize the key features observed for the three FESs and representatively shown for stack~A in Fig.~\ref{fig:BC_shift}: The main consequence of the biased cooling effect is a shift of the field-effect tuned 2DEG density compared to $U_{\text{BC}} =$~\SI{0}{\volt}, the shift increasing with the absolute value of $U_{\text{BC}}$. At the same time, the $U_{\text{TG}}$-tunable $n_{\text{e}}$ range, the capacitive coupling between the TG and the 2DEG and the $\mu$ at each given $n_{\text{e}}$ are unaffected by $U_{\text{BC}}$. Finally, the heterostructures are conductive at room temperature, while this conductivity vanished during the cool-down. At this point it should be highlighted that these observations are identical in all three FESs, although they differ with respect to the presence of a Si cap, the dielectric material and its thickness as well as in the epitaxy method of the heterostructures (CVD vs. MBE), their Ge content and the thicknesses of the SiGe barrier (see Fig.~\ref{fig:heterostructures}). Also FES~A has been fabricated in a semi-industrial process while FES~B and C were fabricated in an academic clean room.

A plausible source for the observed experimental phenomenology is an additional static $U_{\text{BC}}$-dependent electric field which superimposes onto the field resulting from $U_{\text{TG}}$ at the QW at \SI{1.5}{\kelvin}. The charges causing this static field need to be adjustable by the applied $U_{\text{BC}}$ at room temperature, to explain the $U_{\text{BC}}$-dependent shift of the accumulation voltage $U_{\text{Acc}}$. At the same time, to guarantee the observed parallel shift - i.e., constant capacitive coupling - of the $U_{\text{TG}}$ sweeps in Fig.~\ref{fig:BC_shift}\,(a), these charges must be independent of $U_{\text{TG}}$ at \SI{1.5}{\kelvin}.

Since we verified via quantum Hall traces that no parallel conductance channel occurs at \SI{1.5}{\kelvin} after cooling down with $U_{\text{BC}} \neq $~\SI{0}{\volt}, i.e., that transport occurs solely within the 2DEG, these additional charges must be localized. In case these hypothetical charges would build-up in the vicinity of the QW after cooling down with $U_{\text{BC}} \neq $~\SI{0}{\volt}, we would expect a variation in the potential fluctuations affecting the 2DEG in correlation to $U_{\text{BC}}$. As we experimentally find the 2DEG mobility at higher 2DEG densities [Fig.~\ref{fig:BC_shift}\,(b)] and also the minimum 2DEG density $n_{\text{e,min}}$ [Fig.~\ref{fig:BC_shift}\,(a)] to be unaffected by $U_{\text{BC}}$ in all three FESs, we conclude that the charge build-up occurs further away from the QW and is homogeneously distributed. Given that the SiGe barrier is undoped, the most plausible locations are the thin oxidized Si cap and the heterostructure interface with the polycrystalline dielectric oxide. As FES~A, in contrast to B and C, does not contain a (oxidized) Si cap, but nevertheless shows the same behavior under the influence of biased cooling, we exclude the necessity of a Si cap for this effect. Hence, from the previously acquired requirements we conclude that the charges induced at room temperature by the $U_{\text{BC}}$ are localized at the interface between the heterostructure and the polycrystalline dielectric. This interface meets all the criteria. It has been previously shown to host a large enough density of trap states to allow for charge build-ups up to a screening of the capacitive coupling between the TG and the 2DEG \cite{Lu.2011,Wild.2012,Huang.2014,Laroche.2015}. Its location in between the TG and the QW allows the electric field of the trapped interface charges to statically superimpose the electric field of the TG effectively, without being to close to the QW to influence the 2DEG mobility and the minimum 2DEG density. Since all three heterostructures are conductive at room temperature, positive as well as negative $U_{\text{BC}}$ may induce the hypothetical charges exerting the static electric field by loading or unloading trap states at the interface. At the contrary, the freeze-out of the conductance of the heterostructure during the cool-down to \SI{1.5}{\kelvin}, suppresses this loading mechanism of interface trap states.

\subsection{Empirical model for biased cooling of undoped QW heterostructures}

\begin{figure*}
    \includegraphics{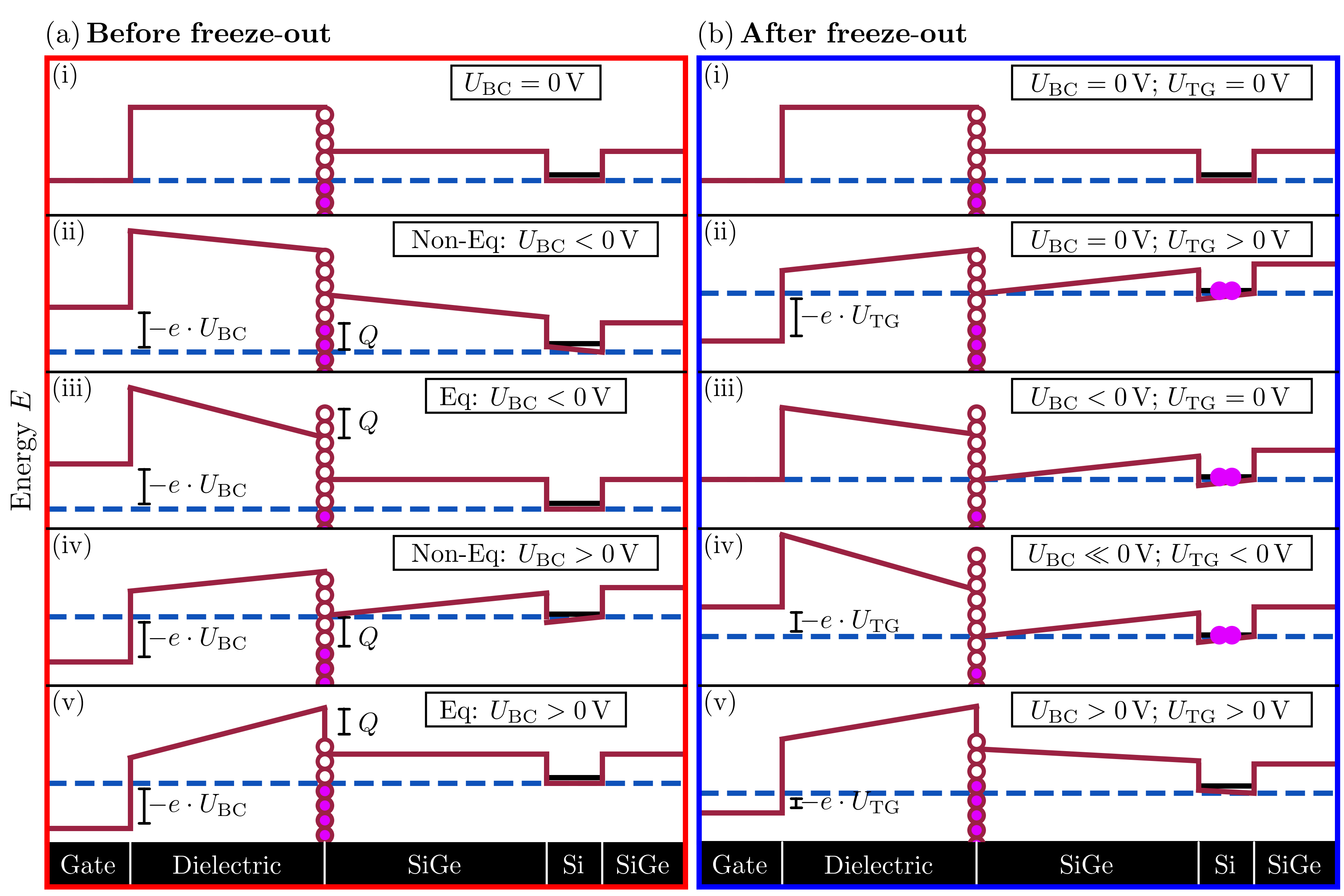}
    \caption{\label{fig:model} Empirical model of the biased cooling effect. The panels (i)-(v) in a) and b) are discussed in detail in the text.
    The dark red line depicts the conduction band edge for different regions of the gate stack indicated at the bottom of the figure. The black line within the Si represents the confined ground state energy of the QW. We chose to fix the lower end of the QW conduction band energy at \SI{0}{\volt}. The blue dashed line illustrates the Fermi energy. The empty red circles between the dielectric and the SiGe depict empty trap states at the dielectric/heterostructure interface. Electrons are shown as magenta colored dots and can fill these states at the interface.
    (a)~Impact of $U_{\text{BC}}$ applied before freeze-out of the heterostructure. At these temperatures, the heterostructure is conductive. Each panel (i) to (v) illustrates a distinct equilibrium or non-equilibrium for specific $U_{\text{BC}}$ situations. The charge $Q$ results from electrons, which are loaded to or unloaded from the interface due to the applied $U_{\text{BC}}$.
    (b)~Impact of $U_{\text{BC}}$ after freeze-out of the conductivity of the heterostructure. At the typical device operation temperature \SI{1.5}{\kelvin}, the electrons at the interface then can not be loaded or unloaded anymore. Each panel (i) to (v) illustrates a specific scenario of combination of $U_{\text{BC}}$ and $U_{\text{TG}}$. Electrons within the Si QW illustrate the accumulation of the 2DEG.
    }
\end{figure*}

Based on this hypothesis, we develop a model for the biased cooling effect in the following. Our model relies on the fact that the top gate and the interface between the heterostructure and the dielectric act like a classical capacitor at room temperature. Essential ingredients of the model are sketched in Fig.~\ref{fig:model}. The mechanism for loading and unloading the interface at room temperature - and hence before the freeze-out of the heterostructure - are sketched in Fig.~\ref{fig:model}\,(a). Panel~\ref{fig:model}\,(a)\,(i) shows the conductance band edge energy of the heterostructure stack (dark red line) for $U_{\text{BC}} =$~\SI{0}{\volt}, with the ground state energy of the Si QW depicted in black. For simplicity, we illustrate this reference case with a flat band edge. These interface states (sketched as red circles) are populated up to the Fermi energy (blue dashed line) with electrons (magenta dots). A non-zero applied $U_{\text{BC}}$ at the TG results in a tilt of the band edge in the sketches. In the case of a $U_{\text{BC}} <$~\SI{0}{\volt} this lifts a certain amount $Q$ of occupied interface states above the Fermi energy (non-equilibrium situation) shown in panel~\ref{fig:model}\,(a)\,(ii). Due to the room temperature conductivity of the FES, these electrons will, however, quickly unload from the interface. This results in a less negative/more positive charge configuration at the interface. As a consequence, two properties of the classical capacitor formed by the top gate and the dielectric/heterostructure interface manifest, as sketched in panel~\ref{fig:model}\,(a)\,(iii): First, the band bending between the top gate and the interface steepens in the dielectric compared to panel~\ref{fig:model}\,(a)\,(ii) proportionally to the charge reconfiguration. Second, outside of the capacitor - i.e., below the interface (in the heterostructure) - there is no electric field. Hence, the conduction band is flat again, as in panel~\ref{fig:model}\,(a)\,(i). In exact analogy, for $U_{\text{BC}} >$~\SI{0}{\volt}, the applied electric field leads to pushing unoccupied interface states containing $Q$ charges below the Fermi energy [see panel~\ref{fig:model}\,(a)\,(iv)] in non-equilibrium. The charge reconfiguration resulting from the room temperature conductivity of the FES hence adds the amount $Q$ of electrons to the interface [see Fig.~\ref{fig:model}\,(a)\,(v)]. Thus, we end up in a more negative charge configuration at the interface compared to panel~\ref{fig:model}\,(a)\,(i). The model in Fig.~\ref{fig:model}\,(a) highlights two features, which are key to explain the experimental observations: The applied $U_{\text{BC}}$ changes the charge state at the interface. Also, the energetic position of the TG relative to the QW ground state energy changes as a function of the applied $U_{\text{BC}}$, while flat band conditions are retained between the interface and the QW. Note that both features result from the fact that the top gate and the dielectric/heterostructure interface behave like the plates of a classical capacitor before freeze-out.

Moving now to temperatures cold enough to freeze-out the conductance of the heterostructure, a major consequence for our model is the suppression of loading or unloading of the trap states at the interface via the heterostructure. Hence, the charges $Q$ trapped at the interface states before cool-down will then be insensitive to the $U_{\text{TG}}$ applied after freeze-out. Figure~\ref{fig:model}\,(b)\,(i), as a reference, shows the scenario of a FES which was cooled down with $U_{\text{BC}} =$~\SI{0}{\volt}, keeping $U_{\text{TG}} =$~\SI{0}{\volt} after freeze-out. The charge density at the interface is non-zero and the QW ground state is above the Fermi energy. Hence no electrons are accumulated in the QW. In order to accumulate electrons within the QW, it is necessary to apply a positive $U_{\text{TG}}$ strong enough to drag the QW ground state below the Fermi energy as illustrated in panel~\ref{fig:model}\,(b)\,(ii). Importantly, since the interface states can not be loaded after freeze-out, the empty interface states pushed below the Fermi level will stay unoccupied, leaving the electric field induced by the electrons trapped at the interface unaffected by $U_{\text{TG}}$ variations. The electron accumulation in the QW is sketched as magenta dots. The density of accumulated electrons in the 2DEG located in the QW is proportional to $U_{\text{TG}}$, experimentally resulting in the typical FES electron density curve shown for $U_{\text{BC}} =$~\SI{0}{\volt} in Fig. \ref{fig:BC_shift}\,(a). Next, Fig.~\ref{fig:model}\,(b)\,(iii) illustrates the case of a negative $U_{\text{BC}}$. Compared to $U_{\text{BC}} =$~\SI{0}{\volt}, the diminished electron density at the interface leads to a less negative electric field superimposing the field created by. This is visible as an additional downwards tilt of the conduction band. The tilt drags the QW ground state closer to the Fermi energy. Now, a less positive $U_{\text{TG}}$ (e.g., $U_{\text{TG}} =$~\SI{0}{\volt}) is required to accumulate electrons within the QW, in line with the electron density curves being shifted towards more negative $U_{\text{TG}}$ in the experiment, as observed in Fig.~\ref{fig:BC_shift}\,(a). For even stronger negative $U_{\text{BC}}$ applied during cool-down, Fig.~\ref{fig:model}\,(b)\,(iv) illustrates the ability to already accumulate electrons in the QW at negative $U_{\text{TG}}$, capturing the experimental observation that the electron density curves are shifted even further towards negative $U_{\text{TG}}$. Figure~\ref{fig:model}\,(b)\,(v) shows the opposite scenario of a positive $U_{\text{BC}}$ applied during the cool-down. The increased electron density at the interface causes a stronger shielding of the $U_{\text{TG}}$ compared to $U_{\text{BC}} =$~\SI{0}{\volt}, resulting in the experimentally observed shift of the electron density curves towards more positive $U_{\text{TG}}$ [see Fig.~\ref{fig:BC_shift}\,(a)].

\begin{figure}
    \centering
    \includegraphics[width=\linewidth]{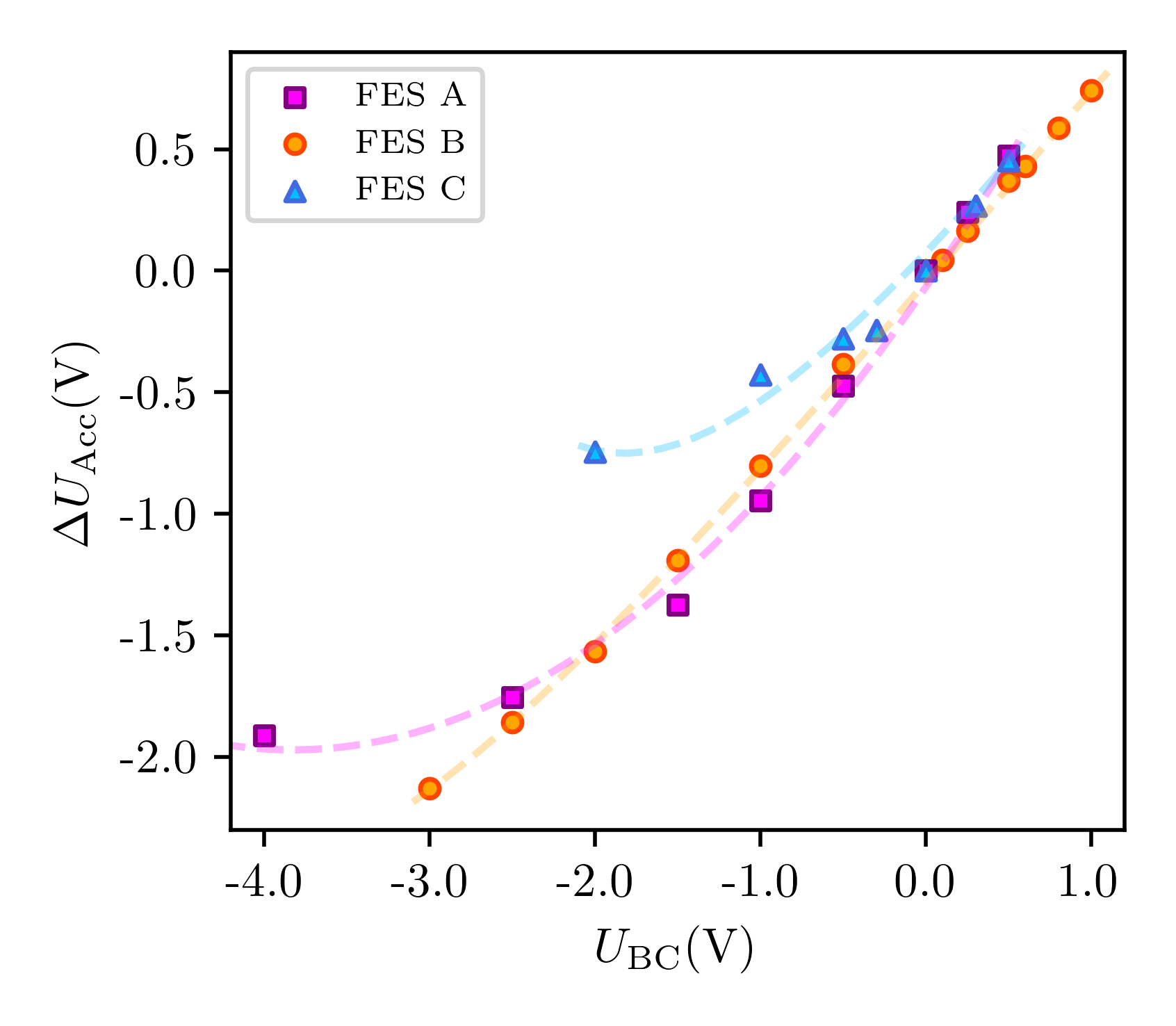}
    \caption{\label{fig:UAcc} Shift of the 2DEG accumulation point $\Delta U_{\text{Acc}}$ as a function of $U_{\text{BC}}$ for all three FESs. The dashed lines are guides to the eyes.}
\end{figure}

Summarizing our experimental observations and the empirical model, applying $U_{\text{BC}}$ at room temperature traps an amount of charges $Q = C_{\text{BC}} \cdot U_{\text{BC}}$ at the dielectric/heterostructure interface [see Fig.~\ref{fig:model}\,(a)\,(ii) to (v)], where $C_{\text{BC}}$ is the capacitive coupling between the TG and the conductive heterostructure during biased cool-down, before freeze-out. After freeze-out, the charge trapping mechanism is suppressed, turning $Q$ into being insensitive to $U_{\text{TG}}$ (applied to the FES at \SI{1.5}{\kelvin}). The constant, static electric field created by the trapped charges $Q(U_{\text{BC}})$ then superimposes the field imposed with a $U_{\text{TG}}$ sweep [see Fig.~\ref{fig:model}\,(b)]. This is equivalent to stating that the accumulation voltage $U_{\text{Acc}}$ of the 2DEG  will be shifted exactly by $U_{\text{BC}}$ with respect to $U_{\text{BC}} =$~\SI{0}{\volt}. As a consequence, our model predicts a linear relationship between $\Delta U_{\text{Acc}} = U_{\text{Acc}} - U_{\text{Acc, $U_{\text{BC}}$ = 0\,\text{V}}}$ and $U_{\text{BC}}$, with a slope $s = 1$. In Fig.~\ref{fig:UAcc}, we test this prediction by displaying $\Delta U_{\text{Acc}}$ for all three FESs. The linear relationship is indeed verified in a significant range of $U_{\text{BC}}$. Also, the slopes $s$ are only slightly smaller than 1, with $s =$~\SI{0.9}{\volt/\volt} for FES~A and C and $s =$~\SI{0.8}{\volt/\volt} for FES~B. This slight deviation seems to indicate that $C_{\text{BC}}$ is a bit smaller than the capacitive coupling at \SI{1.5}{\kelvin}. As a second feature of Fig.~\ref{fig:UAcc}, we observe a deviation from the linear relationship beyond a certain negative value of $U_{\text{BC}}$ for each FES. This suggests that the amount of interface states per energy interval decreases for larger negative $U_{\text{BC}}$ and thus lower energies in Fig.~\ref{fig:model}\,(b). Hence, less charges are unloaded from trap states at the interface at room temperature for these $U_{\text{BC}}$. Note that FES~A does not include a Si cap, while FES~B and C do and that FES~A was produced in a semi-industrial clean room (dielectric SiOx), while FES~B and C were produced in an academic clean room (dielectric AlOx). Both, the deviation from linearity of $\Delta U_{\text{Acc}}$ and the slight variation of the capacitive coupling between \SI{1.5}{\kelvin} and warmer temperatures (slope $s < 1$) thus seem to sensitively depend on non-systematic and subtle details of the FES fabrication.

\subsection{Tunability of the FES operation region after freeze-out}

Up to here, based on the experimental observations, we have concluded that the amount of trapped charges $Q$ at the dielectric/heterostructure interface does not change when varying $U_{\text{TG}} > U_{\text{BC}}$ at \SI{1.5}{\kelvin}. To test this observation in more detail, we investigate whether, after a given cool-down with $U_{\text{BC}}$, the trapped charge density at the interface~$n_{\text{Int}}$ can be influenced at \SI{1.5}{\kelvin} when applying $U_{\text{TG}} = U_{\text{D}}$ \textit{beyond the depletion of the 2DEG} ($U_{\text{D}} < U_{\text{Acc}}$). We denote $U_{\text{Acc, sweep \# = 0}}$ as the accumulation point observed after biased cooling with $U_{\text{BC}}$, tied to its corresponding value of~$n_{\text{Int}}$. We then interpret any deviation of the 2DEG accumulation point observed after applying $U_{\text{D}}$ ($U_{\text{D}} < U_{\text{Acc, sweep \# = 0}}$) as an indicator for a modification of $n_{\text{Int}}$ at \SI{1.5}{\kelvin}. Experimentally, we perform a series of $U_{\text{TG}}$ sweeps at \SI{1.5}{\kelvin}, within the same cool-down. Each sweep $i$ starts from a value $U_{\text{TG}} = U_{\text{D}}$ beyond the depletion of the 2DEG, before the 2DEG is driven into accumulation again, to record the corresponding accumulation point $U_{\text{Acc, sweep \# = i}}$. In the accumulation, we explicitly avoided to enter the saturation regime where tunneling between the 2DEG and the interface sets in. Such sweep series were carried out on FES~A and FES~C, each time for most of the biased cooling voltages tested on that FES (see Fig.~\ref{fig:BC_shift} for FES~A).

\begin{figure}[h!]
    \includegraphics[width=\linewidth]{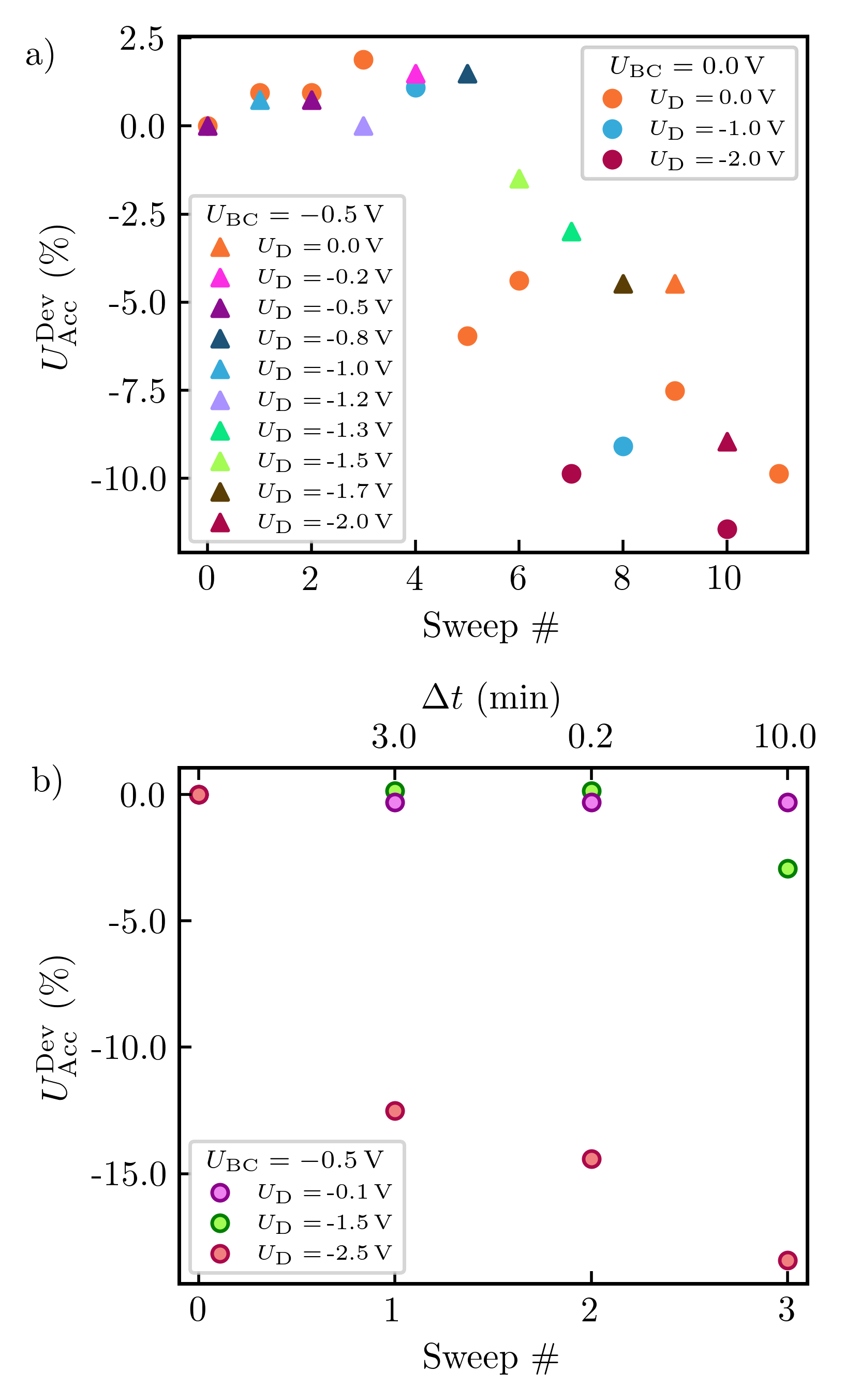}
    \caption{\label{fig:Impact_Interface}~Impact of varying depletion voltages $U_{\text{D}}$ on the accumulation point at \SI{1.5}{\kelvin}. The details of the sweep procedures are discussed in detail in the text.
    (a)~Different accumulation sweeps started from depletion at $U_{\text{D}}$ without thermal cycle of the device, representatively shown for FES~A. The data are shown for two biased cool-downs. The results for $U_{\text{BC}} =$~\SI{0}{\volt} are coded with circles, for $U_{\text{BC}} =$~\SI{-0.5}{\volt} with triangles. We show the relative deviation of the accumulation point $U^{\text{Dev}}_{\text{Acc}}$ [see Eq.~(\ref{eq:U_Dev_Acc})] for each sweep. Each sweep is started from a specific depletion voltage $U_{\text{D}}$, color coded in the figure. 
    (b)~Relative deviation of the accumulation point $U^{\text{Dev}}_{\text{Acc}}$ for different durations $\Delta t$ spent at $U_{\text{D}}$, shown for $U_{\text{BC}} =$~\SI{-0.5}{\volt} on FES~A. The results are shown for three different values of $U_{\text{D}}$, color coded in the figure.}
\end{figure}

For each sweep $i$, we evaluated the relative deviation of the accumulation point of the 2DEG:
\begin{equation}
\label{eq:U_Dev_Acc}
    U^{\text{Dev}}_{\text{Acc}}(\text{sweep \#=i}) = \frac{U_{\text{Acc, sweep \# = i}} - U_{\text{Acc, sweep\# = 0}}}{U_{\text{Acc, sweep \# = 0}} - U_{\text{BC}}}.
\end{equation}
Fig.~\ref{fig:Impact_Interface}\,(a) exemplarily reports observed values of $U^{\text{Dev}}_{\text{Acc}}$ for two separate biased cool-downs of FES~A, $U_{\text{BC}} =$~\SI{0}{\volt} and $U_{\text{BC}} =$~\SI{-0.5}{\volt}. The sweeps are numbered consecutively (sweep~\#) in chronological order starting at zero for the accumulation point obtained directly after biased cool-down at $U_{\text{BC}}$. The chronological series of sweeps $i$ contains random variations of the depletion voltages $U_{\text{D}}$ between sweeps to test the impact of the magnitude of $U_{\text{D}}$ on $n_{\text{Int}}$. Also, some of the $U_{\text{D}}$ values are used twice or more times in one series, to also resolve the role of such repetitions on $n_{\text{Int}}$. For $U_{\text{BC}} =$~\SI{0}{\volt} the series of 11 sweeps contains three different values of $U_{\text{D}}$, randomly varied and repeated. For $U_{\text{BC}} =$~\SI{-0.5}{\volt} the series of 10 sweeps contains a random succession of 10 different $U_{\text{D}}$ values.

For both series - which represent different $U_{\text{BC}}$ and hence two different initial $n_{\text{Int}}$ - we observe that $U_{\text{D}} > U_{\text{BC}}$ tend to shift the 2DEG accumulation point towards more positive $U_{\text{TG}}$ ($U^{\text{Dev}}_{\text{Acc}} > 0$). Equally, $U_{\text{D}} < U_{\text{BC}}$ tend to shift the accumulation point towards lower $U_{\text{TG}}$ ($U^{\text{Dev}}_{\text{Acc}} < 0$). At the same time, the effective action of the applied $U_{\text{D}}$ seems to be statistical. Meaning that applying a certain value of $U_{\text{D}}$ may shift the accumulation point, but will not necessarily do so. For both FES~A and C, we find the magnitude of $U^{\text{Dev}}_{\text{Acc}}$ as well as the probability for a variation of $U^{\text{Dev}}_{\text{Acc}}$ to correlate with the magnitude of $U_{\text{D}}$. Also, there is a clear trend that the more often the FES is subjected to a given value of $U_{\text{D}}$, the higher the probability to observe a non-zero $U^{\text{Dev}}_{\text{Acc}}$ and even a comparatively larger $U^{\text{Dev}}_{\text{Acc}}$. Finally, the data show that the action of successive $U_{\text{D}}$ add up. In Fig.~\ref{fig:Impact_Interface}\,(a), for both $U_{\text{BC}}$, $U^{\text{Dev}}_{\text{Acc}}$ ends up to reach up to -10\,\% after the tenth sweep with mostly negative $U_{\text{D}}$.

We used the same constant $U_{\text{TG}}$ sweep rate in all experiments, meaning that increasing magnitudes of $U_{\text{D}}$ and higher sweep numbers $i$ both imply that the FES spends more time being subjected to voltages $<$~$U_{\text{Acc, sweep \# = 0}}$. To address the influence of this aspect of duration, we performed the additional measurement series Fig.~\ref{fig:Impact_Interface}\,(b) on FES~A, where we vary the duration for which a constant $U_{\text{D}}$ is applied. The cool-down was conducted under $U_{\text{BC}} =$~\SI{-0.5}{\volt} and three values of $U_{\text{D}}$(\SI{-0.1}{\volt}, \SI{-1.5}{\volt}, \SI{-2.5}{\volt}) were tested. For each $U_{\text{D}}$, a series of three sweeps was performed: applying $U_{\text{D}}$ during \SI{3}{\minute} before the sweep~\#\,=\,1, \SI{0.2}{\minute} before sweep~\#\,=\,2 and \SI{10}{\minute} before sweep~\#\,=\,3. For the comparatively small $U_{\text{D}} =$~\SI{-0.1}{\volt}, no observable deviation of the accumulation point is induced, independently of the duration of application of $U_{\text{D}}$. For a stronger $U_{\text{D}} =$~\SI{-1.5}{\volt}, we see that only the longest (\SI{10}{\minute}) and simultaneously last (sweep~\#\,=\,3) exposure to $U_{\text{D}}$ induces a deviation of the 2DEG accumulation point. For the strongest tested $U_{\text{D}} =$~\SI{-2.5}{\volt}, \SI{3}{\minute} (sweep~\#\,=\,1) are sufficient to induce a significant deviation $U^{\text{Dev}}_{\text{Acc}}$. The shorter duration of \SI{0.2}{\minute} in sweep~\#\,=\,2 now also has an impact, slightly larger per time than (sweep~\#\,=\,1).
Compared to the series with $U_{\text{D}} =$~\SI{-0.1}{\volt} and \SI{-1.5}{\volt} this observation may indicate that the statistical rate of the electrostatic process underlying $U^{\text{Dev}}_{\text{Acc}}$ is significantly enhanced for all durations at this larger depletion voltage $U_{\text{D}}=$~\SI{-2.5}{\volt}. The third sweep, which has a duration of \SI{10}{\minute}, further increases $U^{\text{Dev}}_{\text{Acc}}$, but with less impact per time on $U^{\text{Dev}}_{\text{Acc}}$ than the sweeps~\#\,=\,1 and~\#\,=\,2, although sweep~\#\,=\,3 is added to the previous two. We interpret this behavior as an indication for a saturation of the electrostatic effect on $n_{\text{Int}}$, at a given $U_{\text{D}}$.

Summarizing the results of the experiments discussed in Fig.~\ref{fig:Impact_Interface}, we demonstrate that $n_{\text{Int}}$ - which is initialized at room temperature by the choice of $U_{\text{BC}}$ - may be modified during experiments at \SI{1.5}{\kelvin}, by applying a depleting voltage $U_{\text{TG}} = U_{\text{D}}$. At the same time, varying $U_{\text{TG}}$ while the 2DEG is accumulated does not modify $n_{\text{Int}}$. The impact of $U_{\text{D}}$ is strongly statistical. Increasing the duration of $U_{\text{D}}$, repeating its application and also increasing the $U_{\text{D}}$-magnitude increases the probability of the process. In our view, all these experimental signatures strongly hint towards a modification of $n_{\text{Int}}$ via tunneling of electrons between the trap states and the QW.

\section{Conclusion and outlook to applications in quantum circuits}
In conclusion, we have demonstrated that the biased cooling of undoped QW heterostructures creates a static electric field which superimposes on any gate action at the device operation temperature of \SI{1.5}{\kelvin}. While the magnitude of the static electric field scales with the biased cooling voltage $U_{\text{BC}}$ applied at room temperature, it is insensitive to the top gate voltage $U_{\text{TG}}$ action at operation temperature. As a result, the accumulation voltage $U_{\text{Acc}}$ of the 2DEG is reproducibly tunable with $U_{\text{BC}}$, allowing to set $U_{\text{Acc}}$ to be positive as well as negative. The shift of $U_{\text{Acc}}$ depends linearly on $U_{\text{BC}}$ in a wide range. Also, the capacitive coupling of the FES at device operation temperature is not modified by biased cooling. As a consequence, the whole linear $U_{\text{TG}}$-characteristic of the 2DEG density $n_{\text{e}}$ can be shifted deterministically. Importantly, the main measurables of the $U_{\text{TG}}$-tuned 2DEG remain unchanged compared to $U_{\text{BC}}=$~\SI{0}{\volt}: We did not detect any influence of $U_{\text{BC}}$ on the minimal measurable 2DEG density $n_{\text{e,min}}$, on the maximal field-effect tunable density $n_{\text{e,max}}$, nor on the 2DEG mobility within the whole range from $n_{\text{e,min}}$ to $n_{\text{e,max}}$ or on the temporal stability of any chosen $n_{\text{e}}(U_{\text{TG}})$ within this density range.

As we discuss in an empirical model, all our experimental observations are consistent with a charge $Q$\,$=$\,$C_{\text{BC}}$\,$\cdot$\,$U_{\text{BC}}$ being created at the dielectric/heterostructure interface at room temperature via loading or unloading of charge traps. Importantly, the loading and unloading mechanisms are suppressed at the device operation temperature. In addition, the charge is homogeneously distributed, such that it does neither impact $n_{\text{e,min}}$, nor the 2DEG mobility or $n_{\text{e,max}}$. Notably, the mechanism is qualitatively identical, although the three investigated FESs differ (see Tab.~\ref{tab:FES}). In particular the presence of an (oxidized) Si cap at the dielectric/heterostructure interface does not impact the mechanism.

While the trapped charge $Q$ is insensitive to variations of $U_{\text{TG}}$ in the accumulated 2DEG at the device operation temperature of \SI{1.5}{\kelvin}, we have shown that $Q$ can be modified by applying a voltage $U_{\text{D}}$ to the depleted QW. Our experiments indicate that the modification of $Q$ occurs via tunneling between the dieletric/heterostructure interface and the QW.

The ability to shift the operation range of the FES deterministically and reproducibly without affecting the quality features of the 2DEG represents an interesting additional degree of freedom for optimization of gate operation windows. Our results on different heterostructures and our model suggest that the effect should apply to any undoped semiconductor heterostructure. It for example allows to shift "normally on" devices to a  "normally off" operation regime. It has also been shown to allow to avoid initializations of 2DEG in a metastable capacitive coupling \cite{Prager.2021} or leakage regimes in Coulomb blockade devices with integrated charge sensors \cite{Kammerloher.2021}. Applying the biased cooling effect does not require to cycle the device at room temperature. It is sufficient to apply $U_{\text{BC}}$ above the freeze-out temperature of the heterostructure to induce the loading of $Q$ at the interface. 

FES-based quantum circuits with multiple gates, such as current overlapping gate layout designs \cite{Angus.2007,Zajac.2015,Borselli.2015,Zajac.2016} for scalable fault-tolerant \cite{Xue.2022,Noiri.2022,Madzik.2022} and long distance \cite{Struck.2024,Xue.2024,vanRiggelenDoelman.2024} spin qubit operations, require reproducible operation voltages for each gate and are sensitive to large voltage differences applied between neighboring gates during device tuning. The biased cooling effect applied to individual gates in such devices may be advantageous in different aspects, with reduced overhead in tuning, since the device does not need to be thermally cycled to room temperature: The charges trapped at the interface reduce the spatial smearing of the electric field of each gate, which should ease the definition and tunability of few-electron QDs. Furthermore, the pinch-off voltages of gates located in different layers of the overlapping design can be homogenized. Biased cooling also allows to shift the operation window of gates - like tunnel barriers, accumulating plungers or screening gates - responsible for distinct functionalities to similar absolute voltage values. In the current state-of-the-art, very different voltage values typically have to be applied to such gates. The advantage is threefold, reducing the risk of leakage in the device, of local tunneling towards the dielectric/heterostructure interface, with detrimental effects on the operation and charge noise of the device \cite{Massai.2023, Meyer.2023} and also easing the implementation of virtual gates during device tuning. Finally, robustly trapping charges at the interface via biased cooling may represent a route to optimize charge noise of quantum devices, as indications for sweet spots have been observed in preliminary studies \cite{PioroLadriere.2005, Ferrero.2024}.\\

\section*{Acknowledgements}
We acknowledge the financial support of the Deutsche Forschungsgemeinschaft (DFG, German Research Foundation) via Project-ID No. 289786932 (BO 3140/4-2). This work is part of the joint project QUASAR „Halbleiter-Quantenprozessor mit shuttlingbasierter skalierbarer Architektur“ and is supported by the German Federal Ministry of Education and Research (BMBF). The research is also part of the Munich Quantum Valley, which is supported by the Bavarian state government with funds from the Hightech Agenda Bavaria.


%

\end{document}